\documentclass[12pt,preprint]{aastex}
\usepackage{lineno}
\usepackage{times}
\usepackage{natbib}
\linenumbers
\def\F{{\it Fermi}-LAT }
\def\SZ{{\it Suzaku} }

\shorttitle{\SZ Observations of the $\gamma$-ray Lobe in NGC\,6251}
\shortauthors{Takeuchi et al.}

\begin{document}

\title{\SZ X-ray Imaging of the Extended Lobe in the Giant Radio Galaxy NGC\,6251 Associated with the \F Source 2FGL\,J1629.4+8236}

\author{Y.~Takeuchi\altaffilmark{1}, J.~Kataoka\altaffilmark{1}, {\L}.~Stawarz\altaffilmark{2,\,3}, Y.~Takahashi\altaffilmark{1}, K.~Maeda\altaffilmark{1}, T.~Nakamori\altaffilmark{1}, C.~C.~Cheung\altaffilmark{4}, A.~Celotti\altaffilmark{5}, Y.~Tanaka\altaffilmark{2}, and T.~Takahashi\altaffilmark{2}}

\email{uto\_of\_take@suou.waseda.jp}

\altaffiltext{1}{Research Institute for Science and Engineering, Waseda University, 3-4-1 Okubo, Shinjuku, Tokyo, 169-8555, Japan.}
\altaffiltext{2}{Institute of Space and Astronautical Science (ISAS), Japan Aerospace Exploration Agency (JAXA), 3-1-1 Yoshinodai, Chuo-ku, Sagamihara, Kanagawa, 252-5510 Japan}
\altaffiltext{3}{Astronomical Observatory, Jagiellonian University, ul. Orla 171, Krak\'ow 30-244, Poland}
\altaffiltext{4}{National Research Council Research Associate, National Academy of Sciences, Washington, DC 20001, resident at Naval Research Laboratory, Washington, DC 20375, USA}
\altaffiltext{5}{Scuola Internazionale Superiore di Studi Avanzati (SISSA), 34014 Trieste, Italy}

\begin{abstract}

We report the results of a \SZ X-ray imaging study of NGC\,6251, a nearby giant radio galaxy with intermediate FR\,I/II radio properties. Our 
pointing direction was centered on the $\gamma$-ray emission peak recently discovered with \F around the position of the north-west radio lobe 15 arcmin offset from the nucleus.
After subtracting two ``off-source'' pointings adjacent to the radio lobe, and removing possible contaminants in the XIS field of view, we found significant residual X-ray emission most likely diffuse in nature. The spectrum of the excess X-ray emission is well fit by a power law with photon index $\Gamma = 1.90 \pm 0.15$ and a $0.5-8$\,keV flux of $4 \times 10^{-13}$\,erg\,cm$^{-2}$\,s$^{-1}$. We interpret this diffuse X-ray emission component as being due to inverse-Compton up-scattering of the cosmic microwave background photons by ultrarelativistic electrons within the lobe, with only a minor contribution from the beamed emission of the large-scale jet. Utilizing archival radio data for the source, we demonstrate by means of broad-band spectral modeling that the $\gamma$-ray flux of the \F source 2FGL\,J1629.4+8236 may well be accounted for by the high-energy tail of the inverse-Compton continuum of the lobe. 
Thus, this claimed association of $\gamma$-rays from the north-west lobe of NGC\,6251, together with the recent \F imaging of the extended lobes of Centaurus\,A, indicates that particles may be efficiently (re-)accelerated up to ultrarelativistic energies within extended radio lobes of nearby radio galaxies in general. 

\end{abstract}


\section{Introduction}

With the successful launch of the {\it Fermi} Gamma-ray Space Telescope, we now have an unprecedented opportunity to study in detail the $\gamma$-ray emission from different types of astrophysical sources. The 2FGL catalog \citep{2FGL}\footnote{See also the online version at \texttt{http://fermi.gsfc.nasa.gov/ssc/data/access/lat/2yr$\_$catalog/}} contains 1873 sources detected in the 100\,MeV to 100\,GeV range by the Large Area Telescope (LAT) onboard the {\it Fermi} satellite during the first 24 months of the science phase of the mission, which began on 2008 August 4. Among 1297 objects included in 2FGL which are firmly identified or reliably associated with counterparts of known source classes, more than half are blazars, i.e. active galactic nuclei (AGN) with radiative outputs dominated by a beamed jet emission. Non-blazar AGN constitute only a small fraction ($\simeq 3 \%$) of the identified 2FGL population, but nonetheless have already emerged as a new, important, and at the same time relatively diverse class of $\gamma$-ray emitters. The second catalog of AGN detected by \F \citep[2LAC;][]{2LAC} contains ten radio galaxies, four narrow-line Seyfert 1 galaxies, and several other non-blazar-type AGN with prominent starburst components \citep[see also][]{MAGN,FSys}.

Particularly noteworthy in this context is the \F detection of the giant lobes in the nearby radio galaxy Centaurus\,A, extending for about $\sim 5^{\circ}$ in the east-west direction, and $\sim 9^{\circ}$ in the north-south direction \citep[$\sim 300$\,kpc\,$\times~600$\,kpc at the distance of $3.7$\,Mpc;][]{1LAC,FCen}. The \F results regarding the Centaurus\,A lobes have profound implications for the general understanding of the evolution of radio galaxies, their interactions with the surrounding environment, and the production of high-energy particles in the Universe. In particular, prompted by these recent results from the \F and also from the Pierre Auger Observatory \citep{PAO,mos09}, several authors have debated on the high-energy particle acceleration and production of ultra high-energy cosmic rays within the giant lobes of Centaurus\,A and similar systems \citep{har09,osu09,pee11}. 

Besides Centaurus\,A, three other radio galaxies detected so far with \F could be promising targets for investigating the extended $\gamma$-ray emission components associated with radio 
lobes. Prior to the launch of {\it Fermi}, \citet{che07} and \citet{geo08} pointed out that the radio lobes of Fornax\,A might be detected at GeV energies and seen as extended structures 
by the LAT but its associated 2FGL source is faint and currently appears to be a point-source \citep{2FGL}. Similarly, the $\gamma$-ray emission associated with Centaurus\,B and reported recently in 2LAC may be dominated by the lobes, but due to a relatively small angular size of the 
source ($\sim 0.^{\circ} 2$ in the north-south direction) careful analysis of the \F data is needed before drawing any robust conclusions (Katsuta et al. in prep). The giant radio galaxy 
NGC\,6251 already appeared in the first-year \F catalog \citep[1FGL;][]{1FGL} and is another object worth attention due to the large angular extent of its radio structure ($\sim 
1.^{\circ}2$) which in principle could be resolved with \F above 1\,GeV photon energies. 
Based on the one-year of accumulated LAT data, the NGC\,6251 nucleus was included within the 95$\%$ LAT error ellipse of 1FGL\,J1635.4+8228, and its reported $\gamma$-ray flux could be due to the unresolved nuclear jet of the radio galaxy \citep{mig11}. Subsequently however, the position of the 2FGL counterpart of NGC\,6251, namely 2FGL\,J1629.4+8236 (detected at $12\,\sigma$ level), is now shifted north-west with respect to the 1FGL source position, toward the outer jet and radio lobe. We note that previously, the NGC\,6251 galaxy was proposed to be associated with the EGRET source 3EG\,J1621+8203 \citep{muk02}.

NGC\,6251 is a nearby AGN hosting a supermassive black hole with a mass, $M_{\rm BH} \simeq (4 - 8) \times 10^{8} \, M_{\odot}$ \citep{fer99,ho02}. It is classified as an FR\,I radio galaxy \citep{lai83} based on its jet morphology and monochromatic 178\,MHz luminosity of $1.9 \times 10^{24}$\,W\,Hz$^{-1}$\,sr$^{-1}$ \citep{wag77}, although the entire giant radio structure resembles more FR\,II or intermediate FR\,I/FR\,II systems \citep{sch01}. At the distance of $\sim 106$\,Mpc \citep[$z = 0.0247$, conversion scale $0.491$\,kpc/arcsec;][]{weg03} the linear extension of the lobes reads as $\simeq 2.1$\,Mpc, which is a larger physical size than that covered by the outer lobes of Centaurus\,A. The radio jet in NGC\,6251, discussed in detail in \citep{per84}, is likely to be relativistic up to large distances from the core and observed at viewing angles $\theta_{\rm j} \lesssim 40$\,deg, as suggested by the jet-to-counterjet brightness asymmetry on both pc and kpc scales. It can be divided into (i) the bright inner region within the first $\sim 120''$, (ii) a faint central part extending from $\sim 120''$ up to $\sim 180''$, (iii) the \emph{outer region} flaring between $\sim 180''$ and $\sim 270''$, and (iv) a curved and low-surface brightness tail beyond $\sim 270''$. The outer jet region has been detected at X-ray frequencies by {\it ROSAT} \citep{bir93,mac97}, XMM-{\it Newton} and {\it Chandra} \citep{sam04,eva05}. The origin of the detected jet X-ray emission is still an open question and different models have been discussed for such, including thermal emission \citep{mac97}, beamed inverse-Compton radiation \citep{sam04}, and synchrotron process \citep{sam04,eva05}. However, none of the previous X-ray observations pointed to either the north-west or south-east lobe regions, separated by $\sim 0^{\circ}.3$ and $\sim 0^{\circ}.7$ from the NGC\,6251 nucleus, respectively. 

In this paper, we present the results of the newly conducted observations of NGC\,6251 with the \SZ X-ray satellite \citep{mit07} mainly in 2011 April. NGC\,6251 was first observed by \SZ 
in 2010 December \citep{eva11}, but these observations targeted strictly the nucleus of the radio galaxy. This time the pointing direction was intentionally shifted 15 arcmin in the  
north-west direction from the nucleus in order to study the $\gamma$-ray emission peak discovered with \F around the position of the NW radio lobe. In \S\,2, we describe the \SZ observations and data reduction procedure. The results of the analysis are given in \S\,3. The discussion and conclusions are presented in \S\,4 and \S\,5, respectively.

\section{\SZ Observations and Data Reduction}

\SZ is the ideal instrument for the intended study since it provides a low and stable Non X-ray Background (NXB), particularly for diffuse sources \citep{mit07,taw08}. We have therefore observed the NW lobe in NGC\,6251 for 34.2\,ksec (``on'' pointing). In addition, two short observations (18.8\,ksec and 12.1\,ksec) were made in the vicinity of the radio lobe in order to estimate the background flux (``off'' pointings). Hereafter, we denote these three observations as LOBE, BGD1, and BGD2, respectively, and the observation log is provided in Table\,1. Both BGD1 and BGD2 observations were made just after the LOBE observation so that the long-term background fluctuation as well as possible effects caused by the detector degradation can be minimized. In the case of BGD2, an additional short (11.1\,ksec) exposure was obtained in 2011 August, but these data were excluded from the analysis presented here to avoid possible artifacts related to the aforementioned problems. We have checked that the presented results do not change within statistical errors when the additional BGD2 data are included in the spectral analysis.

The observations were made with X-ray Imaging Spectrometer \citep[XIS;][]{koy07} which consists of four CCD cameras each placed in the focal plane of the X-ray Telescope \citep[XRT;][]{ser07}, and with the Hard X-ray Detector \citep[HXD;][]{kok07,tak07}. Each XIS covers a $18' \times 18'$ region on the sky, which is sufficiently large to include the whole NW lobe within a single pointing image. On the other hand, since HXD is a non-imaging instrument with a large field of view $34' \times 34'$ \citep{tak07} and we cannot discriminate between the hard X-ray emission from the nucleus \citep{eva11} and possible emission from the extended NW radio lobe in the HXD data, we only use the imaging and spectral results obtained with XIS in this paper. One of the XIS sensors (XIS 1) has a back-illuminated (BI) CCD, while the other three (XISs 0, 2, and 3) utilize front-illuminated (FI) CCDs. Because of an anomaly in 2006 November, the operation of XIS2 has been terminated and we use only three CCDs. The XIS was operated in the normal full-frame clocking mode with the $3 \times 3$ or $5 \times 5$ editing mode.

In our \SZ pointing, the relatively bright NGC\,6251 nucleus was located well outside the field of view of the XIS. 
However, due to the relatively large point spread function of the XRT onboard \SZ \citep[half power diameter of $\simeq 2.0'$;][]{ser07}, the contamination of the NGC\,6251 nucleus may still, in principle, affect the exposure of the NW radio lobe: the pointing center of our LOBE observation is separated by only $\sim 15'$ from the core. For this reason, we simulated the contamination effect using \texttt{xissim}. In the simulation, we assumed a point source at the position of the nucleus characterized by a power-law continuum with the photon index, $\Gamma = 1.87$, and the normalization $K = 6.79 \times 10^{-4}$\,photons\,keV$^{-1}$\,cm$^{-2}$\,s$^{-1}$ at 1\,keV (corresponding to the $2-10$\,keV unabsorbed luminosity $2.8 \times 10^{42}$\,erg\,s$^{-1}$), as reported in \citep{eva11}. We found that the expected net count rate (the XIS count rate for the ``on'' pointing subtracted by the background count rate estimated from the ``off'' pointings) for the region used in the spectral analysis (see \S\,3.2 below) due to the core contamination is $4.11 \times 10^{-4}$\,s$^{-1}$ only, whilst the total source net count rate is $6.40 \times 10^{-2}$\,s$^{-1}$. Hence, we conclude that the leakage effect is at the level of about $0.64\,\%$, and so that the stray light contamination from the NGC\,6251 nucleus can be considered as negligible when analyzing the diffuse emission of the NW lobe.

We analyzed the screened XIS data, reduced using the \SZ software version 1.2. The screening was based on the following criteria: (1) only {\it ASCA}-grade 0, 2, 3, 4, 6 events were accumulated, while hot and flickering pixels were removed from the XIS image using the \texttt{sisclean} script \citep{day98}, (2) the time interval after the passage of South Atlantic Anomaly was greater than 60\,s, (3) the object was at least $5^{\circ}$ and $20^{\circ}$ above the rim of the Earth (ELV) during night and day, respectively. In addition, we also selected the data with a cutoff rigidity (COR) larger than 6\,GV. In the reduction and the analysis of the \SZ XIS data, HEADAS software version 6.11 and a calibration databases (CALDB) released on 2009 September 25 were used. The XIS cleaned event data-set was obtained in the combined $3 \times 3$ and $5 \times 5$ edit modes using \texttt{xselect}. We note that no spectral variability were observed in the XIS data and the imaging and spectral analyses are discussed in the next section.

\section{X-ray Analysis}

\subsection{Image Analysis}

We extracted the XIS images within the photon energy range of $0.4 - 10$\,keV from only the two FI CCDs (XIS 0, XIS 3) because the BI CCD (XIS1) has lower imaging quality.
In the image analysis, we excluded calibration sources at the corner of the CCD chips. The images of the NXB were obtained from the night Earth data using \texttt{xisnxbgen} \citep{taw08}. Since the exposure times for the original data were different from that of NXB, we calculated the appropriate exposure-corrected original and NXB maps using \texttt{xisexpmapgen} \citep{ish07}. The corrected NXB images were next subtracted from the corrected original images. In addition, we simulated flat sky images using \texttt{xissim} \citep{ish07}, and applied a vignetting correction. All the images obtained with XIS0 and XIS3 were combined and rebinned by a factor of 4 \citep[CCD pixel size $24$\,$\mu$m\,$\times 24$\,$\mu$m, so that $1024 \times 1024$ pixels cover an $18' \times 18'$ region on the sky;][]{koy07}. Throughout these processes, we also performed vignetting correction for all the XIS images for LOBE, BGD1 and BGD2. Finally, the images were smoothed with a Gaussian function with $\sigma = 0.28''$.

The resulting XIS images are shown in Figure\,1, with a 325\,MHz radio contours of NGC\,6251 \citep{WSRT} and the 95$\%$ LAT error circle of 2FGL\,J1629.4+8236 overlaid. When comparing the LOBE, BGD1, and BGD2 images, one can clearly see an excess X-ray emission possibly associated with the NW radio lobe. The apparent enhancements at the edges of the XIS images are artifacts due to an insufficient exposure at the CCD edges caused by small fluctuation of attitude during the observation. Interestingly, at the sensitivity limit of these \SZ observations, no excess X-ray emission was detected from the bright radio hotspot located at the edge of the lobe.
In the context of the study by \citet{har04}, since NGC\,6251 is a low power FR\,I/II source, the X-ray emission from the hotspot is expected to be dominated by the high energy energy tail of the synchrotron continuum rather than by the inverse-Compton component. Specifically, the radio flux of the hotspot of 0.1\,Jy at 1.5\,GHz allows us to estimate roughly the expected X-ray flux by extrapolating the observed hotspot spectrum from lower frequencies as $\sim 2$ nJy. A point source at this flux level is easily detectable by {\it Chandra} or {\it XMM}, but our \SZ non-detection is unsurprising due to the relatively poor angular resolution of the instrument combined with a relatively intense diffuse emission of the surrounding lobe.

\subsection{Spectral Analysis}

In the spectral analysis, we extracted photons from the arc-like source region in the LOBE pointing as shown in Figure\,2. This `arc-like region' we are referring to is a part of a concentric ring centered at the radio core position of NGC\,6251. There we excluded X-ray point sources detected by {\it ROSAT} \citep[see the TABLE1 described in][]{muk02} as background contaminants, assuming the source region radii $2'$ (a typical half-power diameter of the XRT, as described above). We note that the origin of these X-ray features is still under debate, and hence it remains possible that they represent hotspot and enhancement related to the NW lobe itself. In our analysis, however, we removed all such bright X-ray spots, since the main objective of the paper is to detect and to characterize the \emph{diffuse} X-ray emission component associated with the \emph{diffuse} radio structure of the lobe. Similarly, we defined arc-like background regions for both BGD1 and BGD2 pointing using the same detector coordinates as chosen for the LOBE source region. We extracted the spectra from LOBE, BGD1 and BGD2 source regions with \texttt{xselect} for each CCD (XIS 0, XIS 1, XIS 3). Spectral analysis and model fitting were performed with \texttt{xspec} version 12.7.0. In order to improve the statistics, the spectra of XIS0 and XIS3 were summed together using \texttt{mathpha}. Moreover, the spectra of BGD1 and BGD2 were summed for the same reason in the same way. Finally, we made redistribution matrix files (RMFs) and ancillary response files (ARFs) using \texttt{xisrmfgen} and \texttt{xissimarfgen} \citep{ish07}, respectively.

Using these RMFs and ARFs, the corrected spectrum about energy response and the effective area of XIS were obtained. Figure\,3 shows the resulting (background subtracted) spectrum of the LOBE pointing within the energy range $0.4 - 7.5$\,keV. The spectrum could be well fitted by a single power-law continuum with a photon index, $\Gamma = 1.90 \pm 0.15$, moderated by the Galactic absorption only. The Galactic hydrogen column density was fixed as $N_{\rm H} = 5.54 \times 10^{20}$\,cm$^{-2}$ \citep{dic90}. The value of $\chi^{2}$\,/\,d.o.f\,$= 20.54/14$ indicates that this is a satisfactory model for the diffuse X-ray emission component of the NW lobe in NGC\,6251. The details of the fitting results are summarized in Table\,2.
We note that, the spectrum is also equally well reproduced by a thin thermal
plasma models (either bremsstrahlung and/or Raymond-Smith) moderated
by Galactic absorption, with the temperature of $kT$ $\gtrsim$ 3.7 keV. This is rather
high compared with those of thermal plasma of nearby elliptical galaxies
\citep[e.g,][]{mat00}. Moreover, the implied thermal pressure,
$p_{\rm th}$ $\sim$ 2$\times$10$^{-10}$ dyn cm$^{-2}$, is more than four orders of
magnitude larger than the minimum-energy non-thermal pressure of
the lobes (see, section 4.2). We regard these parameters as unrealistic, and
hence we conclude that the diffuse X-ray emission component detected at the position of the NW lobe is purely non-thermal in origin.

\section{Discussion}

The analysis of \SZ data for the NW lobe in NGC\,6251 revealed the presence of X-ray emission, most likely of diffuse nature, well fitted by a power-law continuum with photon index $\Gamma \lesssim 2$ moderated by the Galactic absorption only. The $0.4 - 7.5$\,keV luminosity of the lobe corresponding to the \SZ source extraction region reads as $5.4 \times 10^{41}$\,erg\,s$^{-1}$. The detection of such a non-thermal X-ray emission component is interesting, but not exceptional, since several analogous detections of the X-ray lobes in either FR\,I or FR\,II radio galaxies have been previously reported \citep[see][and references therein]{kat05,cro05,iso11}. The relevance of our \SZ observations of the NW lobe in NGC\,6251 is due to the aforementioned possible (or even likely) association of the lobe with the \F source 2FGL\,J1629.4+8326. Note that in the case of the only radio galaxy for which the lobe emission was robustly associated with a $\gamma$-ray source so far, Centaurus\,A, no diffuse X-ray emission component related to the extended radio structure has been detected \citep[see in this context,][]{har09}. In other words, our observations may potentially provide the very first case of the detected X-ray emission from the large-scale lobe bright in $\gamma$-rays. 

Yet the above statement may be considered as premature, since the 95$\%$ LAT error circle of 2FGL\,J1629.4+8236 includes not only the extended lobe but also the bright `outer jet' region (see \S\,1 and Figure\,1). This region, as mentioned previously, has been resolved in X-rays before with {\it ROSAT}, {\it Chandra}, and XMM-{\it Newton} \citep{mac97,sam04,eva05}. Regardless of the debated origin of the jet-related X-ray photons (synchrotron versus beamed inverse-Compton), large scale jets detected at keV photon energies are in general thought to be the sites of efficient acceleration of high-energy electrons, and as such, to be potential sources of $\gamma$-rays detectable by modern instruments like \F \citep[see, e.g.,][]{sta04,sam04b}. Below we discuss in more detail the possible association of 2FGL\,J1629.4+8236 with the outer jet and with the NW lobe of NGC\,6251, and demonstrate by means of broad-band spectral modeling that the $\gamma$-ray flux can best be accounted for by the high-energy tail of the inverse-Compton continuum of the lobe.

\subsection{Outer Jet}

For the spectral modeling of the outer jet region, we take the 1.4\,GHz flux density, $S_{\rm 1.4\,GHz} = 0.5\pm0.05$\,Jy measured by \citet{sam04} from a VLA map, and assume the radio spectral index $\alpha_r = 0.64\pm 0.05$ within the $1-5$\,GHz range \citep{per84}. We also approximate the X-ray spectrum of this region as a power-law with the photon index $\Gamma = 1.68 \pm 0.13$ within the $0.4-7$\,keV range and the monochromatic flux $S_{\rm 1\,keV} = 4.7 \pm 0.4$\,nJy, as given in \citet{eva05}.\footnote{We note that slightly different X-ray spectral parameters for the outer jet region are claimed by \citet{sam04}, namely $S_{\rm 1\,keV} = 2.3 \pm 0.7$\,nJy and $\Gamma = 1.15 \pm 0.38$ within the $0.5-9$\,keV range. Our re-analysis of archival XMM-{\it Newton} data is however more consistent with that of \citet{eva05}.} We approximate the jet as a cylinder with a radius $R_{\rm j} = 5$\,kpc and \emph{projected} length $\ell_{\rm j} = 40$\,kpc so that the emitting volume is $V_{\rm j} = \pi R_{\rm j}^2 \ell_{\rm j} / \sin \theta_{\rm j}$. 

We model the broad-band emission of the outer jet region in the framework of the beamed inverse-Compton scenario \citep{tav00,cel01}, with the radio emission being due to the synchrotron radiation of non-thermal electrons distributed isotropically within a relativistic outflow, and the X-ray--to--$\gamma$-ray continuum due to Comptonization of the cosmic microwave background (CMB) photons by the same electron population (`beamed EC/CMB' model below). Anisotropy of the CMB radiation in the jet rest frame, as well as Klein-Nishina effects, are properly taken into account \citep{sta05}. For simplicity, we assume homogeneous distributions of the radiating electrons and of the magnetic field within the whole outer jet region, which should be considered as a rough approximation only considering the observed asymmetric radio intensity and polarization transverse profiles of the outer jet \citep{per84}.
 
As shown in Figure\,4 (thin solid line), the $\gamma$-ray flux of 2FGL\,J1629.4+8236 is not well accounted for by the beamed EC/CMB model for the jet with the following `best fit' parameters: the ratio between comoving electron and magnetic field energy densities $\eta_{\rm eq} \equiv U'_e/U'_B = 1.75$, magnetic field intensity $B = 2.3$\,$\mu$G, jet bulk Lorentz factor $\Gamma_{\rm j} =3$, jet viewing angle $\theta_{\rm j} = 17$\,deg, and the electron energy distribution of a power-law form $n'_e(\gamma) \propto \gamma^{-s} \times \exp[-\gamma / \gamma_{\rm max}]$ with $s = 2.2$, $\gamma_{\rm min} = 10$ and $\gamma_{\rm max} = 5 \times 10^5$. With the above model parameters the observed ratio of the inverse-Compton and synchrotron peak luminosities reads as $L_{\rm ic} / L_{\rm syn} \simeq (\delta/\Gamma_{\rm j})^2 \times (U'_{\rm cmb} / U'_{\rm B}) \simeq 25$. Note however that the relatively small jet viewing angle implied by the model, $\theta_{\rm j} = 17$\,deg, may be considered as unlikely considering the large \emph{projected} size of the whole radio structure of NGC\,6251 \citep[see the related discussion in][]{eva05}. On the other hand, the implied jet kinetic luminosities stored in electrons, magnetic field, and cold protons (assuming equal comoving number densities of protons and electrons), namely $L_{\rm e} \simeq 0.7 \times 10^{44}$\,erg\,s$^{-1}$, $L_{\rm B} \simeq 0.4 \times 10^{44}$\,erg\,s$^{-1}$, and $L_{\rm p} \simeq 1.9 \times 10^{45}$\,erg\,s$^{-1}$, giving the total jet power $L_{\rm j}
\simeq 2 \times 10^{45}$\,erg\,s$^{-1}$, could be considered as reasonable values \citep{wil99}. This may imply that if the beamed EC/CMB model for the large-scale jet in NGC\,6251 with the parameters as given above is realistic at all, the outer parts of the outflow may provide some contribution to the observed $\gamma$-ray flux of 2FGL\,J1629.4+8236, at least at the highest photon energies within the LAT range.

\subsection{North-West Lobe}

Next we consider the case of the NW lobe of NGC\,6251 likely associated with the \F source 2FGL\,J1629.4+8236. For the purpose of the spectral modeling we measured radio fluxes for the lobe from published {\it WSRT} maps at 327 and 610 MHz \citep{WSRT}. Additionally, we obtained a new 
VLA\footnote{Calibrated data sets obtained from the NRAO VLA Archive Survey. The National Radio Astronomy Observatory is a facility of the National Science Foundation operated under
cooperative agreement by Associated Universities, Inc.} 1.56 GHz map at 45$''$ resolution with a total 5 hr integration from combining archival 
D-array observations obtained on 1985 December 1, 5, and 1986 Jan 18 (programs AB346 and TEST). Our measurements, $S_{\rm 327\,MHz} 
= 3.10$\,Jy, $S_{\rm 610\,MHz} = 1.75$\,Jy, and $S_{\rm 1.56\,GHz} = 0.907$\,Jy, all assuming $10\%$ errors, were measured using the same source extraction regions as in the \SZ analysis 
(see Figure\,2) \footnote{We note that the radio source extraction region includes a relatively bright radio hotspot mentioned already in section 3.1. By means of careful analysis of the available radio maps we have estimated the contribution of this hotspot to the total radio emission of the lobe as $10\,\%$ at most. i.e. at the level of the assumed radio flux measurement uncertainty}. The lobe also has a higher frequency radio flux measurement by \citet{WSRT}, $S_{\rm 10.55\,GHz} = 213.5 \pm 26.4$\,mJy, and is detected and imaged down to 38 MHz 
\citep{ree90}; due to the low resolution of the latter map, we summed the tabulated fluxes for the 3 components listed west of the central radio source to derive an upper limit to the lobe 
flux, $S_{\rm 38\,MHz} < 39.6$\,Jy. Note that the emerging radio spectral index $\alpha_{\rm 0.326-10.55\,GHz} \simeq 0.8$ is roughly consistent with the X-ray spectral index of the lobe 
measured by \SZ. The lobe is approximated as a sphere with the radius $R_{\ell} = 185$\,kpc, so that the emission region volume is $V_{\ell} = (4/3) \pi R_{\ell}^3$.

For the lobe, we use a similar model as for the outer jet region discussed above, but include in addition extragalactic background light (EBL) photons within the infrared--to--optical range as target photon field for the inverse-Compton scattering (`EC/(CMB+EBL)' model). In the calculations we use the EBL model of \citet{maz07} and account for Klein-Nishina effects. The choice of a particular EBL parameterization does not affect the model results because of the negligible contribution of the EC/EBL emission component to the $\gamma$-ray emission of the lobe within the LAT energy range.
We neglect the contribution of the host galaxy to the target photon field because of the large distance of the analyzed NW lobe from the center ($15' \simeq 450$\,kpc). Finally, we assume homogeneous distributions of the radiating electrons and of the magnetic field within the lobe.

In the framework of the applied simplified model, the $\gamma$-ray flux of 2FGL\,J1629.4+8236 can be accounted for reasonably well (see Figure\,4, thick gray line) by the inverse-Compton emission of the lobe for the following parameters: the equipartition ratio $\eta_{\rm eq} \equiv U_e/U_B = 45$, magnetic field intensity $B = 0.37$\,$\mu$G, and the electron energy distribution of a double-broken power-law form $n_e(\gamma) \propto \gamma^{-s_1}$ for $\gamma_{\rm min} \leq \gamma \leq \gamma_{\rm br}$ and $n_e(\gamma) \propto \gamma^{-s_2} \, \exp[-\gamma / \gamma_{\rm max}]$ for $\gamma > \gamma_{\rm br}$ with $s_1 = 2.0$, $s_2 = 2.5$, $\gamma_{\rm min} = 1$, $\gamma_{\rm br} = 3 \times 10^3$, and $\gamma_{\rm max} = 10^6$. With the above model parameters, the lobe pressure stored in ultrarelativistic electrons and magnetic field is $p_{\rm e+B} \simeq 8.4 \times 10^{-14}$\,dyn\,cm$^{-2}$. The total energy of the structure, being a sum of the work done in displacing a volume $V_{\ell}$ of surrounding gas at pressure $p_{\rm e+B}$, namely $p_{\rm e+B} \, V_{\ell}$, and the energy of the material inside the cavity, $p_{\rm e+B} \, V_{\ell} / (\hat{\gamma}-1)$ (assuming ultrarelativistic equation of state with the adiabatic index $\hat{\gamma} = 4/3$), is then $E_{\rm tot} \simeq 4 \, p_{\rm e+B} \, V_{\ell} \sim 3 \times 10^{59}$\,erg. The lobe's lifetime can be estimated as $t_{\ell} \sim E_{\rm tot} / L_{\rm j} \sim 10 \times \left(L_{\rm j} / 10^{45}\,{\rm erg\,s^{-1}} \right)$\,Myr. We note that any significant proton contribution to the lobe pressure would increase the evaluated values of $E_{\rm tot}$ and $t_{\ell}$.

The EC/(CMB+EBL) model fits the data quite well, and the emerging lobe's parameters seem reasonable. The implied departure from the energy equipartition is rather large, $\eta_{\rm eq} = 
45$, but still within the range $1-100$ established for the lobes in radio galaxies with X-ray measurements alone \citep[e.g.][]{kat05,cro05,iso11}. Furthermore, the expected cooling timescale for the electrons 
emitting the observed $\gtrsim 1$\,GeV photons, i.e., electrons with energies $\gamma \lesssim \gamma_{\rm max} = 10^6$, is roughly $t_{\rm cool} \simeq 2.1 \times \left(\gamma/10^6 
\right)^{-1}$\,Myr. Meanwhile, assuming a typical drift velocity of electrons in the lobe (or expansion velocity of the lobe) $v_{\rm lobe} \lesssim 0.1\,c$, one can infer that these 
$\gamma$-ray emitting electrons can travel only $d \simeq 65 \times \left(v_{\rm lobe} /0.1\,c \right) \times \left(\gamma/10^6 \right)^{-1}$\,kpc before they cool radiatively. This scale 
is much smaller than the extension of the NW lobe in NGC\,6251 ($\sim 0.5$\,Mpc). Interestingly, this finding resembles the case of the giant lobes in Centaurus\,A, and similarly suggests 
efficient in-situ re-acceleration of the radiating particles within the whole volume of the giant lobe \citep[see the discussion in][]{FCen}. Moreover, in both the Centaurus\,A 
and NGC\,6251 radio galaxies, the $\gamma$-ray detections imply that the lobes emit one to two orders of magnitude more energy in $\gamma$-rays than at radio/sub-mm wavelengths.

Finally we note that the non-thermal pressure of the NW lobe in NGC\,6251, $\sim 10^{-13}$\,dyn\,cm$^{-2}$, seems comparable with/bit less than the thermal pressure of the surrounding gaseous medium, $p_{\rm th} \sim (10^{-13}-10^{-12})$\,dyn\,cm$^{-2}$ at the distances $200''-1000''$ from the core. The thermal pressure was estimated by \citet{eva05} based on the extrapolation of the X-ray halo profile associated with the group of galaxies including NGC\,6251 and detected by {\it Chandra} and XMM-{\it Newton} \citep[see also in this context][]{bir93,sam04}. Such an extrapolation may not provide a realistic estimate however, and should rather be considered as an upper limit for the ambient medium pressure. This, together with a likely contribution of (mildly) relativistic protons to the lobe's internal pressure, may suggest that either the extended radio structures in NGC\,6251 are in a pressure equilibrium with the environment, or that these are over-pressured cavities, so that the expansion of the giant lobes in the system is still pressure-driven on Mpc scale (basically beyond the extension of the thermal halo of the surrounding group of galaxies). 

\section{Conclusions}
We have presented a \SZ X-ray observation of the NW lobe in a nearby radio galaxy NGC\,6251. 
Through this observation, we found for the first time non-thermal diffuse X-ray emission associated with the NW lobe.
Since the error circle of the $\gamma$-ray source 2FGL~J1629.4+8236 contains $both$ the NW lobe and outer jet region, 
we discussed the possible origin of $\gamma$-ray emission by detailed modeling of the spectral energy distributions 
assuming a synchrotron plus inverse-Compton emission on CMB background photons.  We argued that, at least at energies 
below 10 GeV, the observed $\gamma$-ray emission is well explained by non-thermal emission from the NW lobe rather 
than from the outer jet region, with reasonable physical parameters (such as magnetic field, electron density etc) 
expected in radio lobes in general. Since the spatial extent of the NGC~6251 radio lobe is quite large ($\sim$ 0.5 Mpc), 
and the expected radiative cooling time is much shorter for $\gamma$-ray emitting electrons, efficient in-situ acceleration 
of radiative particles is necessary within the whole volume of the giant lobe. We also briefly discussed  
the non-thermal pressure stored in ultrarelativistic electrons and magnetic field (not considering a likely contribution of relativistic proton)
of the NW lobe in NGC\,6251, which is estimated as $\sim 10^{-13}$\,dyn\,cm$^{-2}$ from the model, and it seems comparable
with or somewhat less than the thermal pressure of the surrounding gaseous medium.

\acknowledgments

\L .S. is grateful for the support from Polish MNiSW through the grant N-N203-380336. C.C.C.'s work at NRL was supported by NASA DPR S-15633-Y.

\clearpage

\begin{table}
\begin{center}
\caption{{\it Suzaku} XIS observation log.}
\label{tbl-1}
\begin{tabular}{ccccc}
\tableline
Target Name & R.A. [deg] & Dec. [deg] & Exposure [ks] & Obs. Start Time (UT) \\
\tableline
NGC6251\_LOBE & 246.5880 & 82.6370 & 34.2 & 2011 Apr 15 06:37:27 \\
NGC6251\_LOBE\_BGD1 & 247.8200 & 82.9170 & 18.8 & 2011 Apr 16 00:38:47 \\
NGC6251\_LOBE\_BGD2 & 245.4460 & 82.3530 & 12.1 & 2011 Apr 16 10:25:32 \\
\tableline
\end{tabular}
\end{center}
\end{table}

\begin{table}
\caption{Fitting parameters for the power-law model.}
\label{tbl-2}
\begin{tabular}{ll}
\tableline
Parameter & Value \\
\tableline\tableline
$N_{\rm H}$ [cm$^{-2}$] & $5.54 \times 10^{20}$ (fixed) \\
$\Gamma$ & $1.90 \pm 0.15$ \\
$K_{\rm 1\,keV}$ [keV$^{-1}$\,cm$^{-2}$\,s$^{-1}$] & $(9.11 \pm 0.91) \times 10^{-5}$ \\
\tableline
$\chi^2$/d.o.f & $20.54/14$ \\
$P(\chi^2)$ & $1.14 \times 10^{-1}$ \\
\tableline
$F_{\rm 0.4 - 7.5\,keV}$ [erg\,cm$^{-2}$\,s$^{-1}$] & $4.07 \times 10^{-13}$ \\
$L_{\rm 0.4 - 7.5\,keV}$ [erg\,s$^{-1}$] & $5.4 \times 10^{41}$ \\
\tableline
\end{tabular}
\end{table}

\clearpage

\begin{figure}
\begin{center}
\includegraphics[width=150mm]{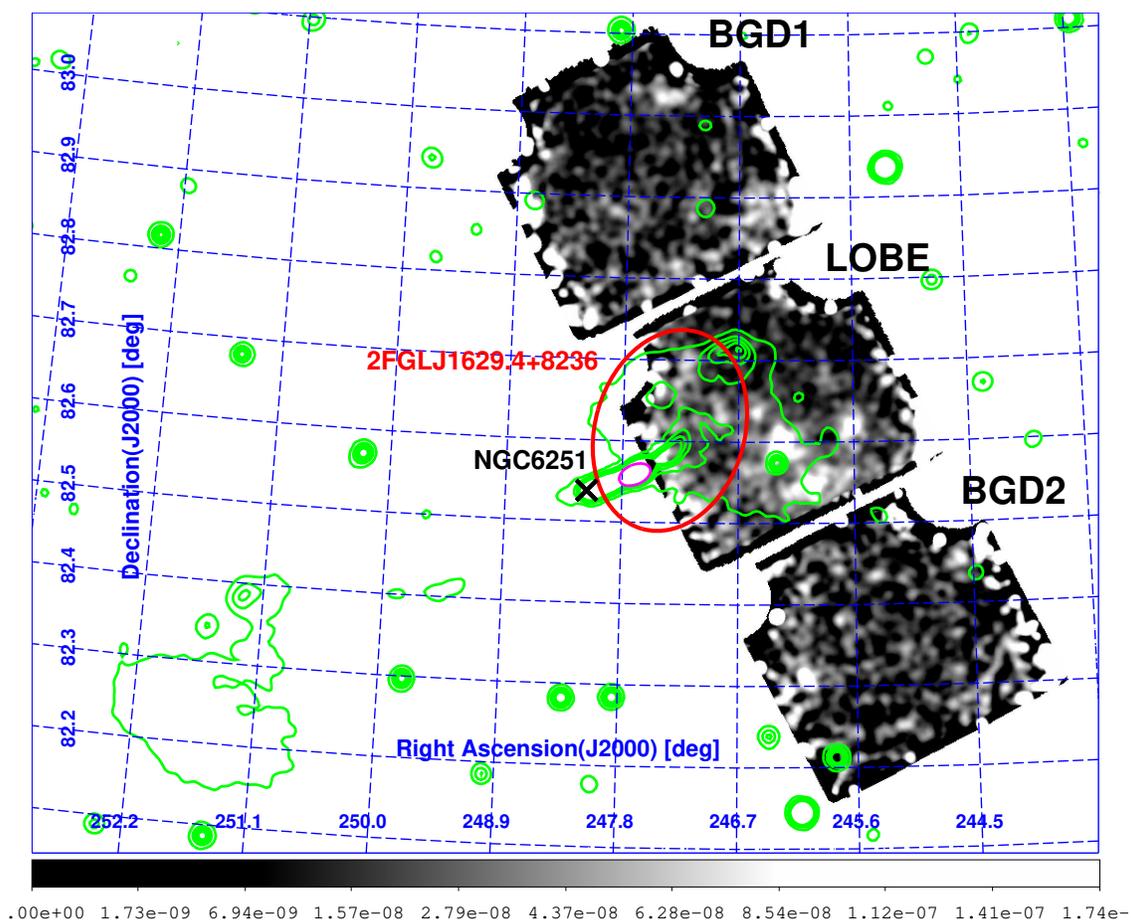}\\
\caption{\SZ X-ray image of the NW lobe in NGC\,6251 (vignetting and exposure corrections applied). Data from XIS 0 and XIS 3 are summed in the $0.4 - 10$\,keV energy band. The image shows the relative excess of smoothed photon counts (arbitrary units indicated in the bottom bar) and displayed with square root scaling. The green contours denote the large-scale radio structure of the source observed by {\it WSRT} at 55\arcsec\ resolution \citep{WSRT} and indicate levels of 8, 31, 54, 77 and 100\,mJy/beam. The red ellipse denotes the 95$\%$ position error of 2FGL\,J1629.4+8236.
The position of the radio core of NGC\,6251 is marked by the black cross at the center and the adjacent `outer jet region' is marked by the magenta ellipse.}
\label{fig-1}
\end{center}
\end{figure}

\begin{figure}
\begin{center}
\includegraphics[width=150mm]{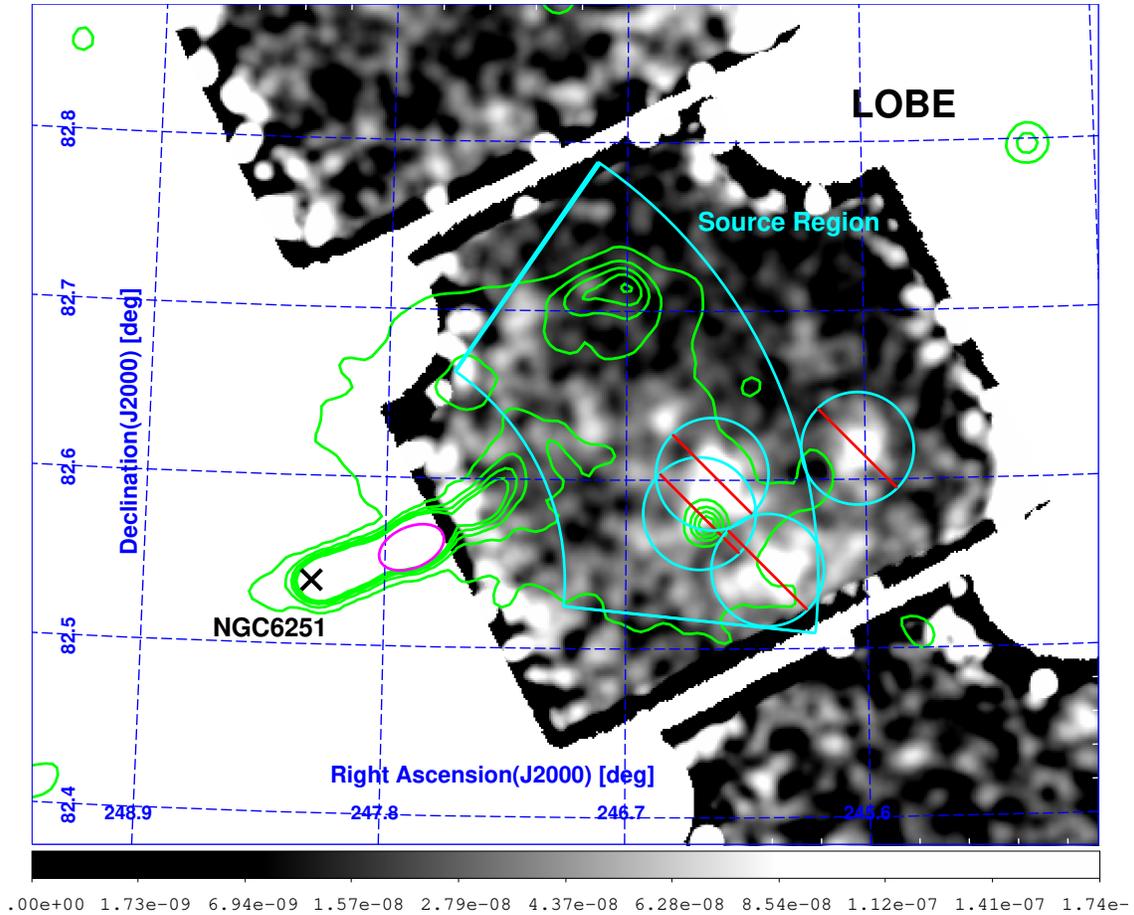}\\
\caption{Same as Figure\,1, but zooming into the NW lobe region. The \SZ source extraction region is denoted by the cyan contours. The point X-ray sources removed from the analysis are marked by cyan circles with red stripes.}
\label{fig-2}
\end{center}
\end{figure}

\clearpage

\begin{figure}
\begin{center}
\includegraphics[width=120mm,angle=-90]{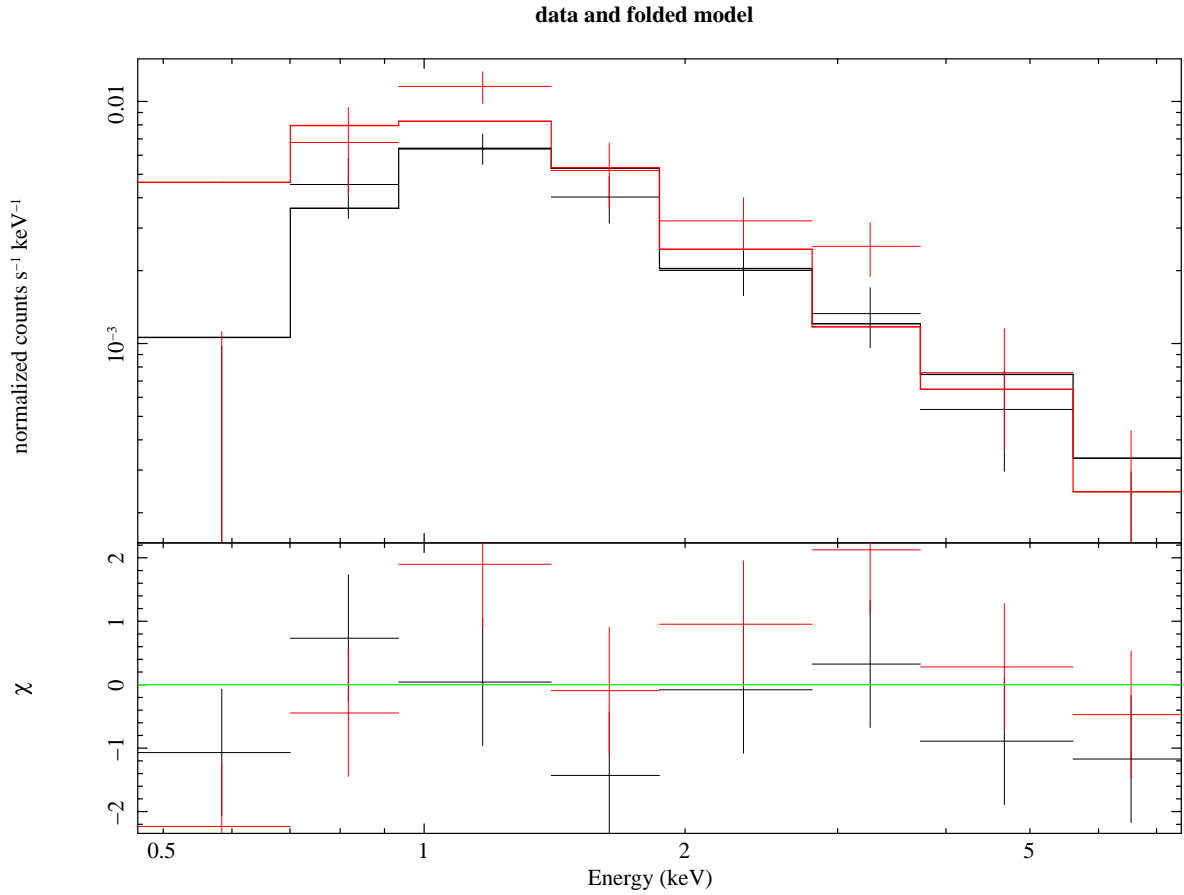}\\
\caption{\SZ XIS spectra of the diffuse emission component coinciding with the NW lobe of NGC\,6251 in the photon energy range $0.4 - 7.5$\,keV, fitted by the model \texttt{wabs $\times$ power-law}. The XIS0 + XIS3 and XIS1 spectra are shown in black and red, respectively.}
\label{fig-3}
\end{center}
\end{figure}

\clearpage

\begin{figure}
\begin{center}
\includegraphics[scale=1.75]{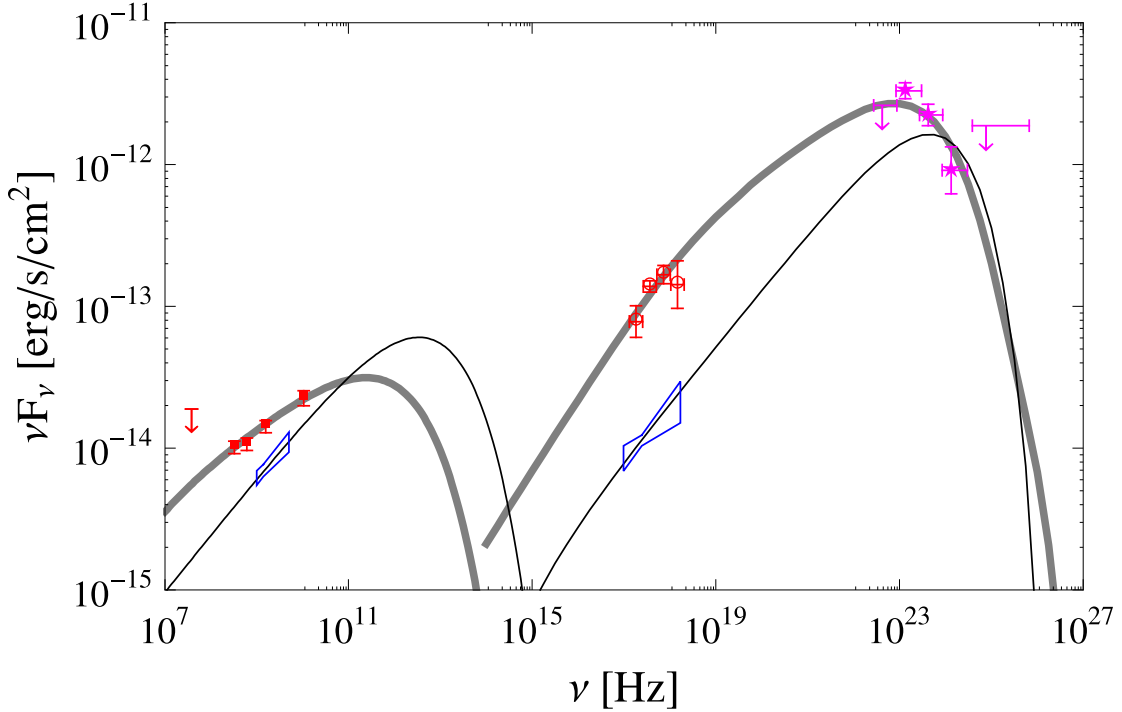}\\
\caption{Broad-band spectral energy distributions of the outer jet and the NW lobe in NGC\,6251, including the \F source 2FGL\,J1629.4+8326. The radio and X-ray data for the outer jet (blue bow-ties) are from \citep[based on the maps presented in Perley et al. 1984]{sam04}, and \citet{eva05}, respectively (see \S\,4.1). The radio fluxes of the NW lobe (red squares and red arrow) are described in section~4.2. The X-ray fluxes for the lobe, as found in this paper, are given in the $0.5 - 1.0$\,keV, $1.0 -2.0$\,keV, $2.0 - 4.0$\,keV, and $4.0 - 8.0$\,keV bins (open red circles). Finally, the LAT fluxes of 2FGL\,J1629.4+8326 (magenta stars) are taken from the 2FGL catalog \citep{2FGL}. Thin solid line represents the beamed IC/CMB model for the outer jet (see \S\,4.1), assuming the association with 2FGL\,J1629.4+8326. Thick gray line represent the IC/(CMB+EBL) model for the NW lobe (see \S\,4.2), again assuming the association with 2FGL\,J1629.4+8326.}
\label{fig-4}
\end{center}
\end{figure}

\end{document}